\documentclass[aps,prl,twocolumn,10pt]{revtex4}

\usepackage[]{graphicx}
\usepackage{amsmath}
\usepackage[T1]{fontenc}
\usepackage{mathptmx}

\begin{document}

\title{Weak continuous monitoring of a flux qubit using coplanar waveguide resonator}

\author{G. Oelsner}
 \author{S. H. W. van der Ploeg}
 \author{P. Macha}
 \author{U. H\"ubner}
 \author{D. Born}
 \author{S. Anders}
 \author{E. Il'ichev}
 \author{H.-G. Meyer}
 \affiliation{Institute of Photonic Technology, PO Box 100239, D-07702 Jena, Germany}
 \author{M. Grajcar}
 \affiliation{Department of Solid State Physics,
Comenius University, SK-84248 Bratislava, Slovakia.}
\author{S. W\"unsch}
\author{M. Siegel}
\affiliation{Karlsruhe Institute of Technology, Institut f\"ur Mikro- und Nanoelektronische Systeme,
Hertzstra\ss e 16, D-76187 Karlsruhe, Germany}
\author{A.N. Omelyanchouk}
\affiliation{B. Verkin Institute for Low Temperature Physics and Engineering, 47 Lenin Avenue, 61103 Kharkov, Ukraine}
\author{O. Astafiev}
\affiliation{NEC Nano Electronics Research Laboratories. Tsukuba, Ibaraki, 305-8501, Japan}

\date{\today}

\begin{abstract}
We study a flux qubit in a coplanar waveguide resonator by measuring transmission through the system. In our system with the flux qubit decoupled galvanically from the resonator, the intermediate coupling regime is achieved. In this regime dispersive readout is possible with weak backaction on the qubit. The detailed  theoretical analysis and simulations give a good agreement with the experimental data and allow to make the qubit characterization.

\end{abstract}

\maketitle


Implementation of a scalable solid state quantum system for either quantum information processing devices
or quantum limited detectors requires high-fidelity read-out methods for its quantum states. For superconducting
qubits the switching of a $dc$ SQUID or a probe junction to the normal state was used for the determination of the
qubit states \cite{Nak99, Wal00, Vion02}. However, the switching to the normal state
leads to strong  backaction from the measurement device on the qubit.
In order to minimize detector-qubit backaction the dispersive
type of measurements has been proposed \cite{Green_a02, Green_b02} and implemented \cite{Ili03, Born2004}.
In the frame of this method a qubit is coupled to a resonator with a resonance frequency
much lower than the qubit transition frequency.
The phase of the oscillator modes was monitored to perform qubit measurements.

If the qubit's characteristic frequency is comparable to the eigenfrequency of the resonator,
coherent interaction between the quantum oscillator and the artificial atom (qubit) can be realized.
In the optical domain this is known as cavity quantum electrodynamics (cQED) which
can precisely describe the entanglement of a single two-level
quantum system (e.g. spin 1/2) and a cavity field \cite{Rai01}.
Solid state cQED based on superconducting circuits has been suggested \cite{Ble04} and demonstrated \cite{Wal04,Abdufarrukh2008} using superconducting qubits coupled to a
superconducting transmission line
resonator. To observe resonator-qubit entanglement the coherent coupling strength between them should
exceed the rate of incoherent processes in this system. Quantitatively this means that
$g>\kappa;\gamma$ where $\hbar g$ is the coupling energy between qubit and resonator, and
$\kappa$ is the resonator loss rate and $\gamma$ is the qubit dephasing rate
(so-called strong coupling regime).

Besides studying the fundamentally interesting strong coupling regime, the same architecture can be used for the readout of qubits as well \cite{Bianchetti2009}. In the dispersive regime, when the qubit transition frequency is far from the resonator frequency, even a quantum non-demolition (QND) measurement can be performed provided the photon numbers in the resonator are low enough. The phase change in the resonator at a particular (probe) frequency
can be monitored to perform qubit state readout \cite{Gra04}. This measurement can be done in either
a weak continuous \cite{Gra04, Wal04} or pulse regime \cite{Bianchetti2009}.

In the dispersive regime, when the resonator acts as a qubit detector, the resonator-qubit backaction
can be minimized by decreasing the coupling  $g$. However, in order to measure
resonator-qubit dynamics the condition $g>\kappa$ should be fulfilled in order
to avoid the performance degradation of the resonator as a detector.
In this work, we test experimentally the resonator-qubit system with \emph{intermediate}
coupling where  $g > \kappa$ but $g \sim\gamma$.

\begin{figure}[tbp]
\includegraphics[width=8cm]{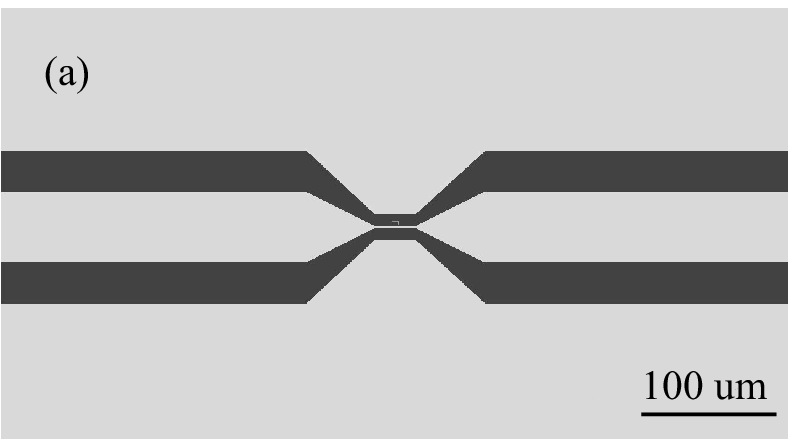}
\includegraphics[width=8cm]{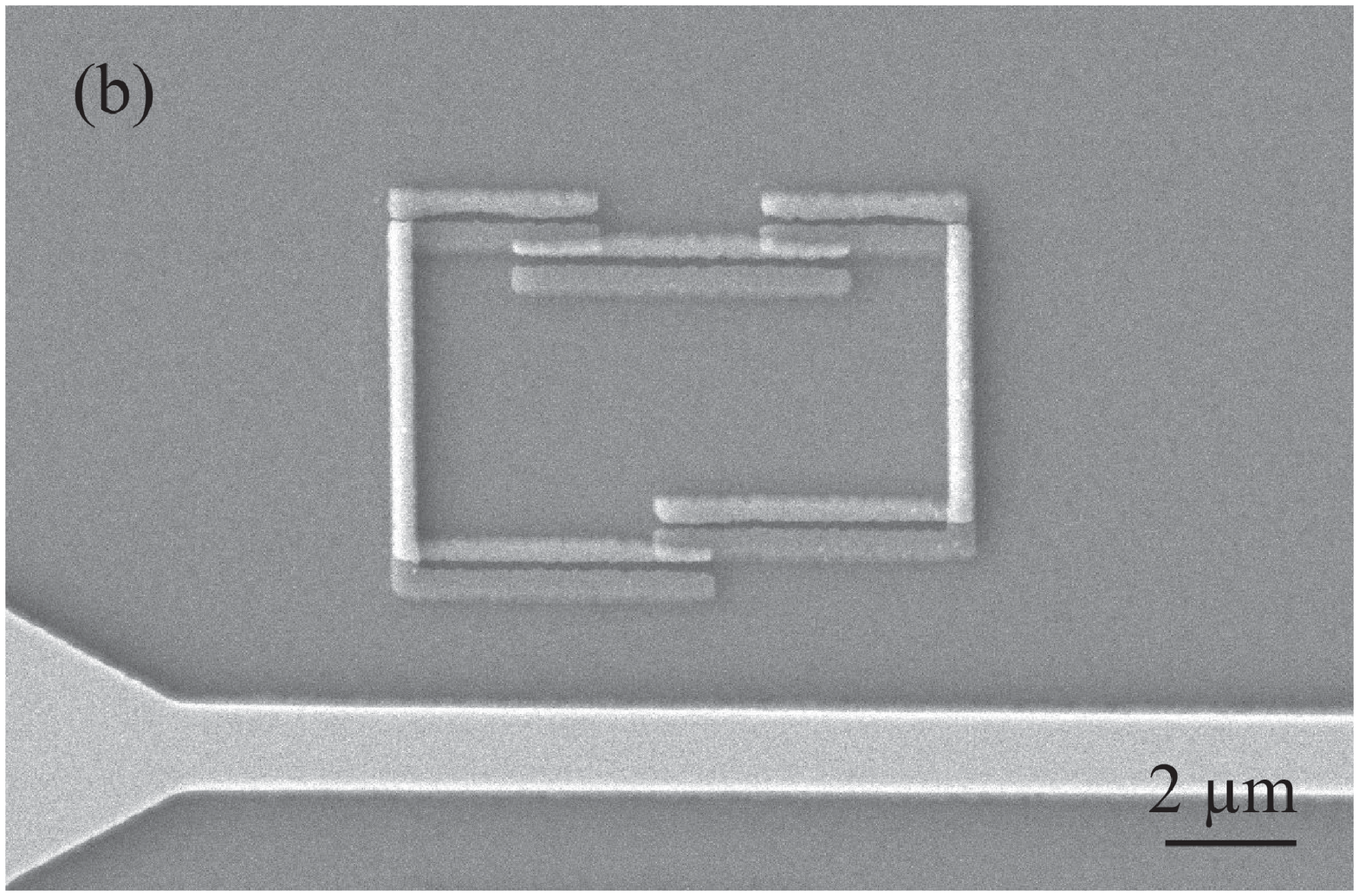}
\caption{(a) A micrograph of the cental part of the resonator. The central wire and the ground plane are tapered. (b) Electron micrograph of the qubit and the central line of the resonator\label{fig:design}}
 \end{figure}

The resonator was fabricated by e-beam lithography and CF$_{\text{4}}$ reactive-ion
etching of a 200~nm thick Nb-film deposited on an undoped silicon substrate. The
length of the resonator's central conductor is $L$ = 23~mm, its width 50~$\mu$m and
the gap between the central conductor and the ground plane 30~$\mu$m, which results in a wave impedance of about $Z$ = 50 $\Omega$ and a resonance
frequency $\omega_r/2\pi \approx$ 2.5~GHz for the fundamental, half-wavelength, mode.
The resonator is coupled to the in- and output waveguides via two identical
gaps of 90 $\mu$m in the central conductor corresponding to a coupling capacitance $C_\kappa \approx$ 6~fF. At temperatures
$T < \text{50 mK}$, the resonator quality factor $Q = \omega_r / \kappa \approx 2\cdot10^4$, corresponding to $\kappa/2\pi \approx$ 0.13~MHz determined by loading loss in the external 50 $\Omega$ impedance.
In the middle of the resonator, the central conductor is tapered to a
width of 1~$\mu$m and a length of 30~$\mu$m with 9~$\mu$m gap (see
Fig.~\ref{fig:design}(a)), which provides the necessary qubit-resonator coupling.

The Al persistent current qubit is fabricated in the central (tapered) part
of the resonator by using conventional two-angle shadow evaporation technique. It
is galvanically decoupled from the resonator line (Fig.~\ref{fig:design}(b)), making the qubit fabrication independent on the resonator and providing flexibility in design and fabrication of more complicated circuits. The qubit loop (size of $2.5\times5~\mu\text{m}^{\text{2}}$) is inductively coupled to the resonator. The coupling constant $g/2\pi\text{ = 3 MHz}$ has been numerically estimated
and is of the same order as the decay rates of flux qubits
\cite{Wal04, Yos06}. The loop is interrupted by three Josephson
junctions. Two of them have a nominal size of $800 \times 180$~nm$^{\text{2}}$, while
the third is about $35 ~\%$ smaller.

Two cryoperm shields and one superconducting (lead) shield enclosed the sample
in order to minimize the influence of external magnetic fields. The sample was
thermally anchored to the mixing chamber of a dilution refrigerator, providing
a base temperature of less than $10\text{ mK}$. Measurements of the
amplitude  as well as the phase $\varphi$ of the transmitted
signal through the qubit-resonator system as a function of applied power and
frequency were performed by using a network analyzer. Attenuators are located at
different temperature stages on the input line preventing thermal
noise of those stages to reach the sample. The transmitted signal is amplified by a cryogenic
amplifier placed at the 4.2~K stage and several room temperature amplifiers. An
isolator (circulator with one terminated port) placed at the mixing chamber is used to protect the sample from backaction of the amplifier. The qubit energy bias is controlled via two small external coils located underneath
and above the sample.

The qubit-resonator system in the natural (flux) basis is described by the Hamiltonian \cite{Hauss2008}
\begin{equation}
	 H = \frac{1}{2}\epsilon(\Phi_x)\sigma_z-\frac{1}{2}\Delta\sigma_x+\hbar\omega_ra^\dagger a+
	\hbar g\sigma_z(a+a^\dagger)\; ,
	\label{Eq:H}
\end{equation}
where $\epsilon(\Phi_x)=2\Phi_0I_p(1-\Phi_x/\Phi_0)$ is the bias of the flux
qubit controlled by the external $dc$ magnetic flux $\Phi_x$, $\Phi_0=h/2e$ is the magnetic flux
quantum and $I_p$ is persistent current of the flux qubit $\sigma_{x,z}$ are the Pauli matrices,
$\Delta$ is the tunnel splitting of the flux qubit,   and
$a^\dagger$, $a$ are the photon creation and annihilation operators.
After transformation to the eigenbasis of the qubit and neglecting the small
diagonal terms, the Hamiltonian can be rewritten as
\begin{equation}
	H \approx \frac{\hbar\omega_q}{2}\sigma_z+\hbar\omega_ra^\dagger a+\hbar g_\epsilon(a^\dagger + a)\sigma_x\; ,
	\label{Eq:Hr}
\end{equation}
where $g_\epsilon = g\cdot\Delta/\sqrt{\epsilon^2+\Delta^2}$ is the normalized coupling, and
$\hbar\omega_q=\sqrt{\epsilon^2+\Delta^2}$ is the energy level separation of the qubit.

For $n$ photons in the resonator, the qubit-resonator dynamics
is completely confined to
a two dimensional subspace with basis  $|g\rangle|n\rangle$ and $|e\rangle|n-1\rangle$.
The eigenenergies relative to the ground state of the Hamiltonian of Eq. (\ref{Eq:Hr}) can be expressed in the analytical form \cite{Ble04}
\begin{equation}
\frac{\Delta E_\pm}{\hbar} =\left(n+\frac{1}{2}\right)\omega_r+\frac{\omega_q}{2}\pm
\sqrt{\left(\frac{\delta}{2}\right)^2+{g_\epsilon}^2(n+1)}\; ,
	\label{Eq:E}
\end{equation}
were $\delta = \omega_{q}-\omega_{r}$ is the qubit-resonator detuning. By measuring the transmission through the
resonator, the transition frequencies of the qubit-resonator system can be probed.
If there is no interaction between the qubit and the resonator, the system should respond only at
frequencies corresponding to the eigenenergies of the qubit $\hbar\omega_q$ and resonator
$\hbar\omega_{r}$.

 \begin{figure}
\includegraphics[width=8cm]{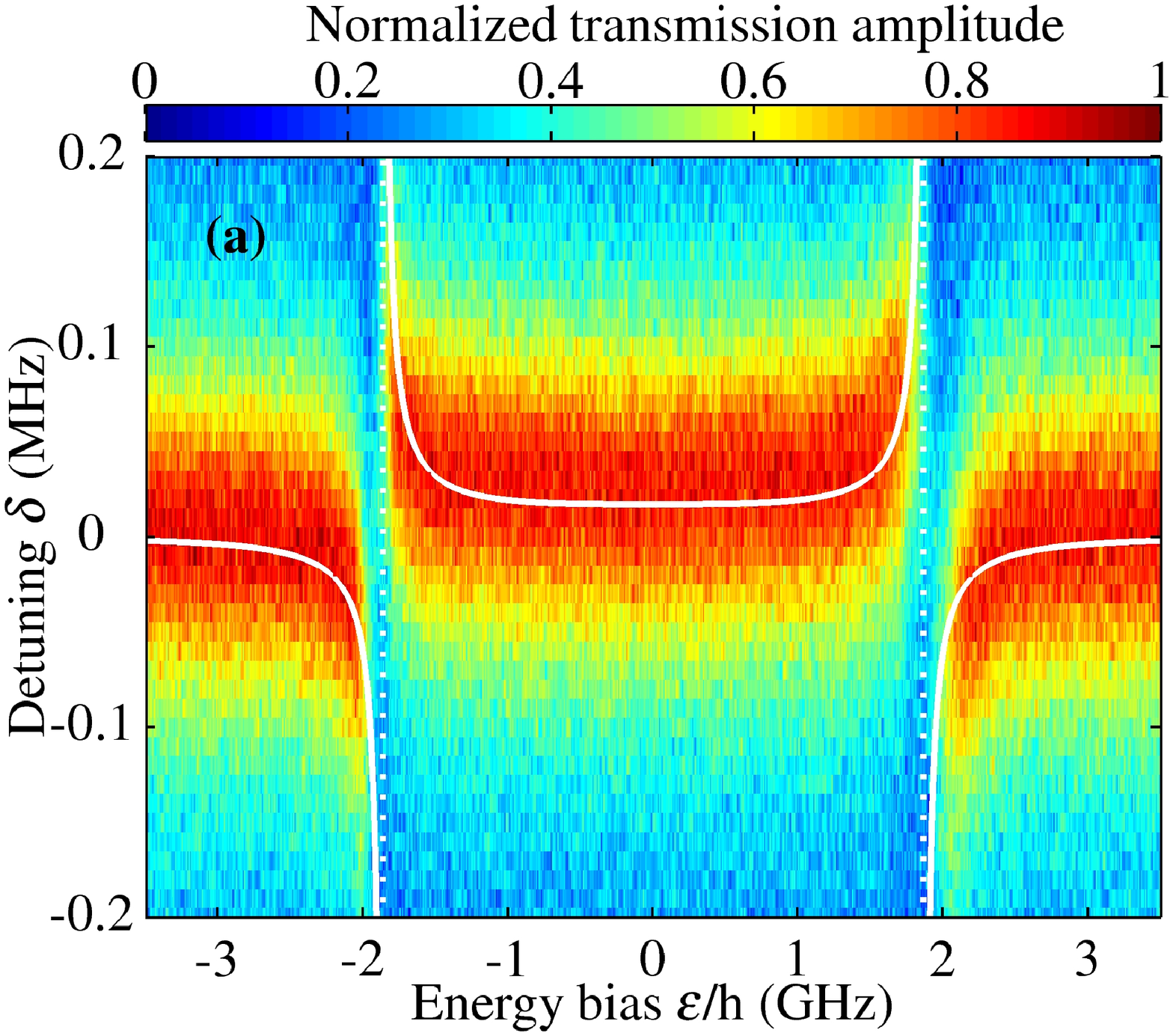}\\
\includegraphics[width=8cm]{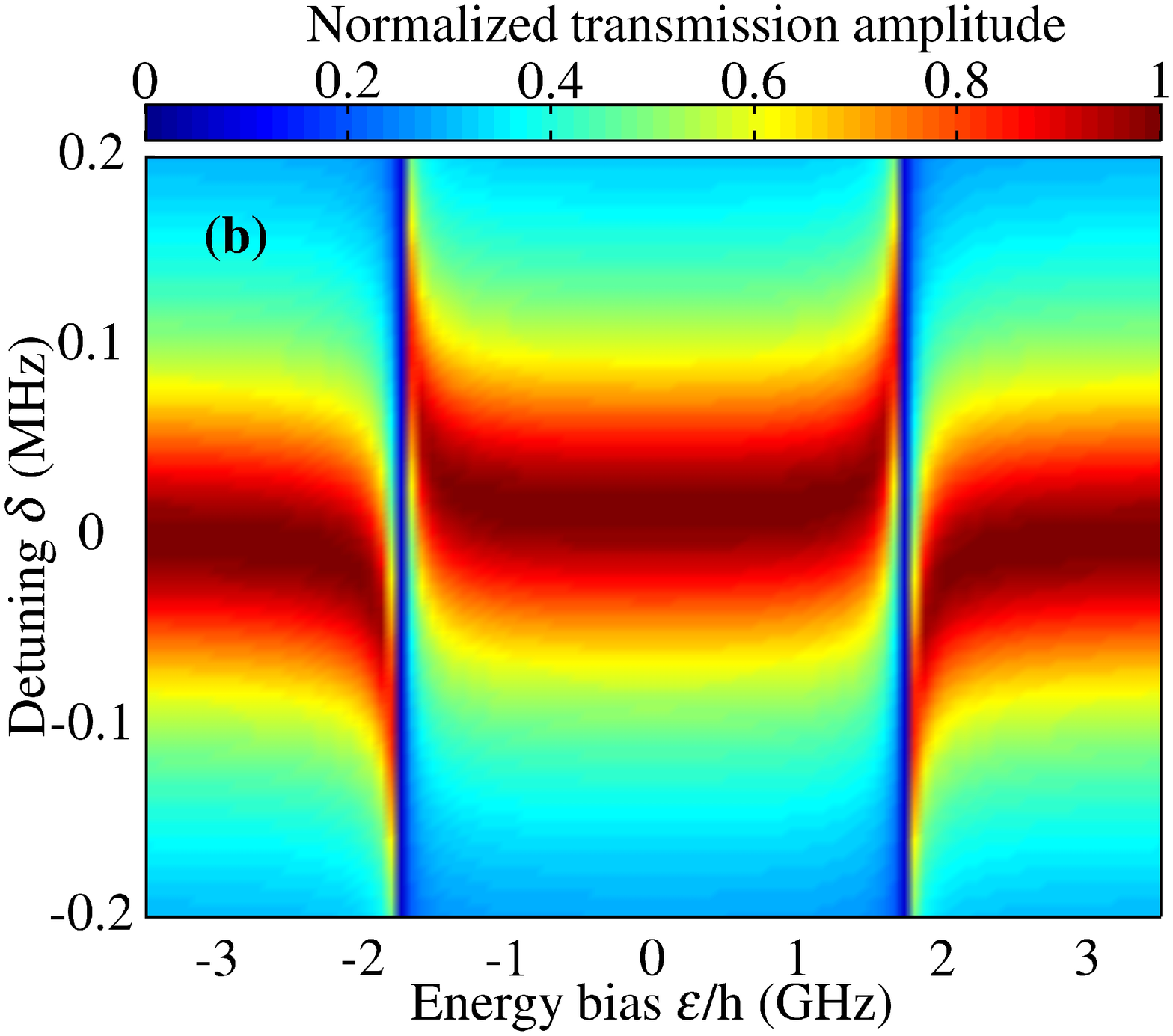}%
\caption{
(Color online) a) Normalized transmission amplitude of the resonator as a function of the qubit energy bias $\epsilon$ and the driving frequency.
The data were measured at a nominal temperature of the mixing chamber below 10 mK, ensured
$k_BT\ll\hbar\omega_q, \hbar\omega_r$.
The calculated frequencies of the lowest photon transition in the
qubit-resonator system are depicted as white solid lines. b) Theoretical calculations. The number
of photons is taken to be less than one, other parameters are taken from the experiment. $\gamma$ is adjusted for better correspondence of the transmission in vicinity of anticrossings.}
\label{fig:pic2}
\end{figure}

Fig.~\ref{fig:pic2}(a) demonstrates the experimentally measured transmission amplitude $t$. The finite coupling $g$ allows to observe anticrossings. The crossing points are visible at $\omega_q=\omega_r$
(dotted white lines in Fig.~\ref{fig:pic2}(a)).
The white solid lines were calculated using Eq. (\ref{Eq:E}) for the lowest photon transition, which corresponds to the mean photon number in the resonator less then one.

In contrast to strong coupling where two well-resolved spectral lines separated by the
vacuum Rabi frequency at the resonance point ($\delta = 0$) have been observed \cite{Wal04}, we have detected the broadening of the transmission peak only and the peaks disappeared at $\delta = 0$. Numerical simulations show that this is a feature of the intermediate coupling when indeed $g\lesssim\gamma$. However, in spite of the fact that the coupling is relatively weak,
we have clearly observed the anticrossing features.

To investigate the qubit-resonator dissipative dynamics in details, we analyze the Markovian master
equation for the density matrix $\rho$
\begin{equation}
\dot{\rho}=-\frac{i}{\hbar}\left[  H,\rho\right]  +L[\rho],
  \label{Eq:Master}
\end{equation}
where the dissipative (Lindblad) term $L = L_r + L_q$ presents two incoherent processes: dissipation in the resonator (photon decay) \begin{equation}\label{4}
    L_r = \frac{\kappa}{2} (2a\rho a^\dagger - a^\dagger a \rho -  \rho a^\dagger a)
\end{equation}
and qubit decoherence: relaxation with rate $\gamma_1$ and pure dephasing with rate $\gamma_\phi$
\begin{equation}\label{5}
    L_q = \frac{\gamma_1}{2} (2\sigma^-\rho \sigma^\dagger - \sigma^\dagger \sigma^- \rho -  \rho \sigma^\dagger\sigma^-) + \frac{\gamma_\phi}{2} (\sigma_z \rho \sigma_z - \rho),
\end{equation}
where $\sigma_\pm = (\sigma_x \pm i\sigma_y$)/2.

We consider a 1D (coplanar waveguide) resonator of size $L$ ($-L/2 \leq x \leq L/2$) driven by the external voltage $V_{in}(t) = V_{in}\cos\omega t$ from the left hand side through the pointlike coupling capacitance $C_\kappa$ located at $x = -L/2$. The excited voltage field within the resonator can be presented as $V(x, t) = - V_r (a e^{-i\omega t} + a^\dag e^{i\omega t})\sin(\pi x/L)$, where $V_r = \sqrt{\hbar\omega/C_r}$ and $C_r$ is the resonator central line capacitance. The interaction Hamiltonian is $H_{int} = C_\kappa V_{in}(t)V(-L/2,t)$ and the total driven system Hamiltonian in the rotating wave approximation in vicinity of the qubit-resonator resonance $\omega_r \sim \omega_q$ and with small detunings $\delta_r = \omega - \omega_r$ and $\delta_q = \omega - \omega_q$ is

\begin{align}
H \approx -\frac{\hbar\delta_q}{2} \sigma_{z} -\hbar\delta_r a^{\dag}a+
\hbar g_\epsilon(a^\dagger \sigma_- + a\sigma_+) + \frac{\hbar\Omega}{2}(a^{\dag}+a),
\end{align}
where $\hbar\Omega = C_\kappa V_{in} V_r$.

Solving the stationary master equation Eq. (\ref{Eq:Master}) ($\dot{\rho} = 0$), one can calculate the observable quantities. In particular, the expectation value of the photon field in the resonator in the considered weak driving limit (with the truncated the photon space to two states) is
\begin{equation}
\langle a\rangle = -\frac{i\Omega/2}{\delta'_r - g_\epsilon^2/\delta'_q}\;,
\end{equation}
where $\delta'_r = \delta_r + i\kappa/2$, $\delta'_q = \delta_q + i\gamma$ are redefined detunings and $\gamma = \gamma_{1}/2+\gamma_{\phi}$ is the total qubit dephasing. The amplitude of the field at the output of the resonator (at $x = L/2 + 0$) is $V_{out}(t) \approx - i\omega C_\kappa Z V_r (\langle a\rangle e^{-i\omega t} + \langle a^\dag \rangle e^{i\omega t})$. And using the explicit expression for the photon relaxation rate $\kappa \approx 2 Z C_\kappa^2\omega^2/C_r$, we find the transmission amplitude (defined as $t = V_{out}/V_{in}$) to be
\begin{equation}
t = -\frac{i \kappa/2}{\delta'_r - g_\epsilon^2/\delta'_q}\;.
\label{Eq:trans}
\end{equation}

In order to characterize the qubit energy, we carry out spectroscopic measurements in the following way. Two waves are applied: an excitation wave, which excites the qubit transitions and thereby results in a change of the resonator transmission, and a probing one close to $\omega_r$, which enables to measure the transmission. Using the qubit parameters, $t$ is calculated according to Eq. (\ref{Eq:trans}) and plotted in Fig.~\ref{fig:pic2}(b) as a density plot versus $\delta_r$ and $\epsilon$. In this plot, $\gamma$ was used as an adjustable parameter for better correspondence with the experimental data in Fig.~\ref{fig:pic2}(a) and found to be $\gamma/2\pi$ = 0.26 MHz.

\begin{figure}
\includegraphics[width=8cm]{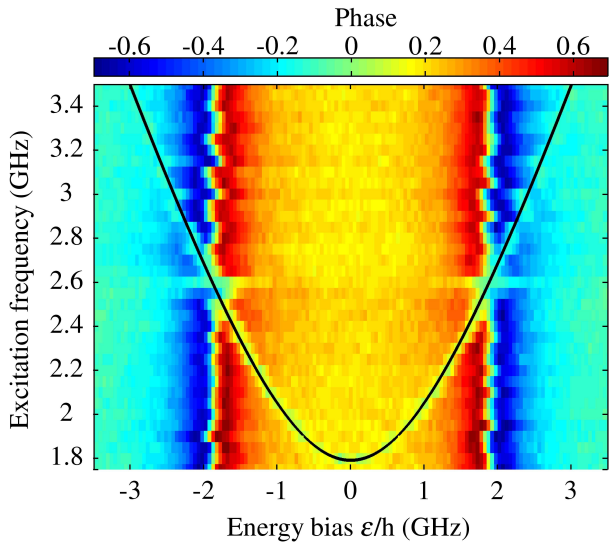}%
\caption{ (Color online)
Transmission phase of the resonator as a function of the qubit excitation frequency $\epsilon/h$. The black line depicts the fitted energy level splitting of the qubit.}
\label{fig:pic3}
\end{figure}

To understand the effect of the qubit state on the resonator transmission far away
from the anticrossings ($\left|\delta \right| \gg g_\epsilon$), the Hamiltonian in Eq. (\ref{Eq:Hr}) can be transformed \cite{Ble04} to
\begin{equation}
H \approx \hbar\left( \omega_r + \frac{{g_\epsilon}^2}{\delta} \sigma_z \right) a^\dagger a + \frac{\hbar}{2}\left( \omega_q + \frac{{g_\epsilon}^2}{\delta} \right) \sigma_z \; .
\label{Eq:Hr3}
\end{equation}
The first term contains the ac Zeeman shift of the resonance frequency, which depends on the state of the qubit as well as on its detuning.
The $ac$-Zeeman frequency shift is visualized by the color profile in Fig.~\ref{fig:pic3}. It is positive (negative) for $\delta < 0$ ($\delta > 0$). If the driving microwave field is in resonance
with the qubit energy levels, the expectation value of $\sigma_z$ becomes zero, resulting
in disappearance of the ac Zeeman shift. Therefore, the green traces appear in Fig.~\ref{fig:pic3}
following the relation for energy levels separation of the qubit $\omega_q(\epsilon)$. By fitting
this curve (black solid line) the persistent current $I_p= 180$ nA as well as the gap
$\Delta/h=1.8$ GHz of the qubit were determined. A similar effect was observed by
Schuster \emph{et al.} with a charge qubit \cite{Schuster05}.

In conclusion, we demonstrate that in the galvanically decoupled resonator-flux qubit system, an intermediate coupling, i.e. $g \lesssim \gamma$, is possible. With the intermediate coupling, the main quantum mechanical features of this system are observable: the $ac$-Zeeman shift, the anticrossing between the single photon in the resonator and the qubit, as well as the level splitting of the qubit versus external bias.

We gratefully acknowledge the financial support of the EU through the EuroSQIP and Solid project and the financial support by the DFG IL150/b-1.
EI acknowledges the financial support
from the Federal Agency on Science and Innovations of the Russian
Federation under contract N 02.740.11.5067.
M.G. was partially supported by the Slovak Scientific Grant Agency
Grant No. 1/0096/08, the Slovak Research and Development Agency under the
contract No. APVV-0432-07 and No. VVCE-0058-07, ERDF OP R\&D,
Project CE QUTE ITMS 262401022, and via CE SAS QUTE.

\end{document}